\documentclass[aps,prd,showpacs,preprintnumbers]{revtex4}

\usepackage{graphics}
\usepackage{color}
\usepackage{multirow}
\usepackage{amsmath}
\newcommand{\beq}{\begin{equation}}
\newcommand{\eeq}{\end{equation}}
\newcommand{\beqa}{\begin{eqnarray}}
\newcommand{\eeqa}{\end{eqnarray}}
\newcommand{\re}{{\rm Real\/}\,}

\newcommand{\tr}{{\rm tr\/}\,}
\newcommand{\om}{\omega}
\newcommand{\rw}{\rho(\omega)}
\newcommand{\rwo}{\rho_1(\omega)}
\newcommand{\rwt}{\rho_2(\omega)}
\newcommand{\gt}{G_E(\tau)}

\newcommand{\tc}{T_c}
\newcommand{\kt}{\kappa/T^3}
\newcommand{\nt}{N_\tau}
\newcommand{\alms}{\alpha^{\overline{MS}}_s(\mu)}
\definecolor{red}{rgb}{1.0,0.0,0.25}

\def \etal{{\sl et al.\/}}
\def \jhep{{\sl J.\ H.\ E.\ P.\ }}

\def \np{{\sl Nucl.\ Phys.\/}}
\def \pl{{\sl Phys.\ Lett.\/}}
\def \pr{{\sl Phys.\ Rev.\/}}
\def \prl{{\sl Phys.\ Rev.\ Lett.\/}}

\begin{document}
\title{Heavy Quark Momentum Diffusion Coefficient from Lattice QCD}
\author{Debasish Banerjee}
\email{debasish@theory.tifr.res.in}
\affiliation{Department of Theoretical Physics, Tata Institute of Fundamental
         Research,\\ Homi Bhabha Road, Mumbai 400005, India.}
\author{Saumen Datta}
\email{saumen@theory.tifr.res.in}
\affiliation{Department of Theoretical Physics, Tata Institute of Fundamental
         Research,\\ Homi Bhabha Road, Mumbai 400005, India.}
\author{Rajiv Gavai}
\email{gavai@tifr.res.in}
\affiliation{Department of Theoretical Physics, Tata Institute of Fundamental
         Research,\\ Homi Bhabha Road, Mumbai 400005, India.}
\author{Pushan Majumdar}
\email{tppm@iacs.res.in}
\affiliation{Department of Theoretical Physics, Indian Association for the 
Cultivation of Science,\\
 Kolkata 700032, India.}

\begin{abstract}
The momentum diffusion coefficient for heavy quarks is studied in a deconfined
gluon plasma in the static approximation by investigating a correlation
function of the color electric field using Monte Carlo techniques.  The
diffusion coefficient is extracted from the long distance behavior of such a
correlator.  For temperatures $ \tc < T \lesssim 2 \tc$, our nonperturbative
estimate of the diffusion coefficient is found to be very different from the
leading order perturbation theory, and is in the right ballpark to explain the
heavy quark flow seen by PHENIX at RHIC.
\preprint{TIFR/TH/11-41}
\end{abstract}
\maketitle

\section{Introduction}
\label{sec.intro}

The charm and the bottom quarks are very
important tools in our quest to understand the nature of the
quark-gluon plasma created in the relativistic heavy ion collision
experiments. Since the masses of both of them are much larger than
the temperatures attained in RHIC, and in LHC, one expects these
quarks to be produced largely in the early pre-equilibrated state of
the collision, and thus provide a window to look into the early stages of
the fireball. Furthermore, perturbative arguments suggest that the energy
loss mechanism for energetic heavy quarks in medium should be different
from that of the light quarks. A comparative study of the energy
loss for the heavy and light quark jets therefore leads to crucial
insights into the way the quark-gluon plasma interacts.

For light quark jets, gluon radiation (``bremsstrahlung'') is expected
to be the leading mechanism for energy loss in medium \cite{bdmps}. It
has been argued that gluon bremsstrahlung is suppressed for jets of
heavy quarks \cite{dk}, and collisional energy loss may be the
dominant mechanism for thermalization of not-too-energetic heavy quark
jets \cite{mt,mustafa}. Since collision with a thermal quark does not
change the energy of a heavy quark substantially, one would expect
that the thermalization time of the heavy quarks is much larger than
that of the light quarks. As most of the elliptic flow is developed
early, the azimuthal anisotropy parameter, $v_2$, of the hadrons with heavy
quarks can be expected to be much less than that of the light hadrons.

Interesting predictions follow from these simple, weak coupling-based
intuitions, which can be, and have been, checked in the RHIC
experiments. One expects a mass ordering of the elliptic flow:
$v_2^h \ \gg \ v_2^D \ \gg \ v_2^B$.  Here $h, D, B$ refer to the
light hadrons, mesons of the $D$ family (one charm and one light
quark) and those in the $B$ family (one bottom and one light quark).
The nuclear suppression factor, $R_{AA}$, is also expected to show a
hierarchy: $R_{AA}^h \ \ll \ R_{AA}^D \ \ll \ R_{AA}^B$.
Experimentally, on the other hand, it was found that the heavy flavor 
mesons show a large elliptic flow, $v_2^D \lesssim v_2^h,$ and a
strong nuclear suppression, $R_{AA}^D \gtrsim R_{AA}^h$, the nuclear
suppression being comparable to that of $\pi^0$ for $p_T \ > \ 2$ GeV
\cite{exp,exp2}.

Even if the kinetic energy of the heavy quark is $\mathcal{O}(T)$,
where $T$ is the temperature of the fireball, its momentum will be
much larger than the temperature. It is, therefore, changed very
little in a single collision, and successive collisions can be treated
as uncorrelated. Based on this picture, a Langevin description of the
motion of the heavy quark in the medium has been proposed
\cite{svetitsky,mt,mustafa}. $v_2$, the elliptic flow parameter, can
then be calculated in terms of the diffusion coefficient of the heavy
quark in the medium. The diffusion coefficient has been calculated in
perturbation theory \cite{svetitsky,mt}. While the experimental
results for the elliptic flow of the charmed mesons and its $p_T$
dependence seem to be well described by this formalism for moderate
$p_T \lesssim 2$ GeV, the value of the diffusion coefficient 
needed to explain the
experimental data is found to be at least an order of magnitude lower
than the leading order (LO) perturbation theory (PT) result
\cite{exp2} \footnote{Note that the observed flow depends not only on
  the diffusion coefficient but also on various other details like the
  geometry of the collision, evolution, equation of state of the
  plasma, etc. See \cite{rapp} for a comprehensive review.}  Furthermore,
the LOPT value itself leads to heavy flavor flow which falls way short of 
the data.  The contribution of the next-to-leading order (NLO) in perturbation 
theory has been calculated recently \cite{cm}. Although, it was found to 
change the LO result by a large factor at temperatures $\lesssim 2
\tc$, this should perhaps be taken as an indication of the inadequacy of
perturbation theory in obtaining a reliable estimate for the diffusion 
coefficient in the temperature range of interest.  

A nonperturbative estimate of the diffusion coefficient, $D$, in QCD
is, therefore, essential to understand the heavy quark flow in the
Langevin formalism.  Lattice QCD, together
with numerical Monte Carlo techniques, provides the only way of doing
first principle nonperturbative calculations in the quark-gluon
plasma. Unfortunately, such calculations are done in Euclidean space,
and extracting a real time object like the diffusion coefficient
requires an analytic continuation, which is extremely
difficult. However, estimation of various transport coefficients have
already been attempted, with varying degrees of success
\cite{rtreview}. In order to estimate the heavy quark diffusion
coefficient, one can study the correlator of the heavy quark current,
$\bar{Q} \gamma^i Q$. Early attempts to extract $D$ this way showed
that the correlator has very little sensitivity to $D$ \cite{pt}. 
Preliminary results of a calculation of $D$ extracted from $\bar{Q}
\gamma^i Q$ correlator, using much finer lattices than used before, have been
presented recently  in the temperature range between 1.5 --- 3
$\tc$ for a gluon plasma \cite{qm11}. They do find a value which is much
lower than the LOPT result, and in the right ballpark to explain the
experimental heavy quark flow.

Two of the difficulties in extracting the diffusion coefficient from
$\bar{Q} \gamma^i Q$ correlator are: i) the behavior of the structure
of the spectral function near the $\om \sim 2 m_Q$ regime can affect
the structure at low $\om$, and ii) the diffusion coefficient is
obtained from the width of the narrow transport peak at $\om \to 0$,
which is difficult to extract. In the infinitely heavy quark limit,
another approach to the diffusion coefficient has been suggested in
Refs. \cite{ct,clm}.  In this static limit, the propagation of heavy
quarks is replaced by Wilson lines, and the formalism of
Ref. \cite{mt} reduces to the evaluation of retarded correlator of
electric fields connected by Wilson lines \cite{ct}.  In
Ref. \cite{ct}, this formalism was used to calculate the diffusion
coefficient for the $\mathcal{N}$ = 4, SU($N_c \to \infty$) gauge
theory, using the AdS/CFT correspondence. A parametric dependence on
coupling very different from weak coupling perturbation theory was
obtained.  On the other hand, for the pure SU(3) gauge theory, a
leading order perturbative calculation of this correlator led to a
negative value for the diffusion coefficient at moderate temperatures
\cite{bllm}.

The formalism outlined in Ref. \cite{clm} is suitable for Monte Carlo
calculation on the lattice. As we outline in the next section, this
involves the calculation of Matsubara correlators of color electric
field operators, and extracting the low frequency part of the spectral
function from it. Of course, the usual problems of extraction of the
spectral function from the Matsubara correlator mean that calculation
of the diffusion coefficient remains a highly nontrivial task. A first
attempt to calculate the diffusion coefficient from the electric field
correlator lead to very large values of the diffusion coefficient,
close to the perturbation theory value \cite{meyer}. Preliminary
results from a recent calculation \cite{langelage}, on the other hand,
gave values in the temperature range 1.5-3 $\tc$ close to the
experimental results.

In this work, we use the formalism of \cite{ct,clm} to calculate the
diffusion coefficient of the deconfined gluonic plasma in a moderate
temperature range, $\tc < T \lesssim 2 \tc$. The aim is to understand
whether a small diffusion coefficient, as found in the analysis of the
experimental data \cite{exp2}, is consistent with QCD. The plan of the
paper is as follows. In the next section we outline the formalism. In
Sec.  \ref{sec.details} we explain the operators and the algorithm.
Sec. \ref{sec.main} has our results. A discussion of the results,
including their connection with experiments, is contained in
Sec. \ref{sec.summary}. Some details of Secs. \ref{sec.details} and
\ref{sec.main} are relegated to the appendices.

\section{Formalism}
\label{sec.theory}

In this section, we outline the formalism of Refs. \cite{ct,clm} in
more detail. We first sketch the arguments leading to the Langevin
formalism, and then discuss the quantum field theoretic calculation of
suitable correlation functions. This discussion closely follows
Refs. \cite{mt, ct, clm}. Then we discuss the issues related to the
extraction of the diffusion coefficient from the Matsubara correlator.

It is easy to see why the motion of a quark much heavier than the
system temperature can be described in the Langevin formalism. If the
kinetic energy is $\sim T$, then the momentum, $\sim \sqrt{M T}$, is
not changed substantially in individual collisions with thermal gluons
and quarks, which can only lead to a momentum transfer $\sim
T$. Therefore, the motion of the heavy quark is similar to a Brownian
motion, and the force on it can be written as the sum of a drag term
and a ``white noise'' , corresponding to uncorrelated random
collisions: \\ 
\beq 
\frac{d p_i}{dt} \ = \ - \eta_D p_i \ +
\ \xi_i(t), \qquad \langle \xi_i(t) \xi_j(t^\prime) \rangle \ = \ \kappa
\ \delta_{i j} \ \delta(t-t^\prime) .
\label{eq.langevin} \eeq
From Eq. (\ref{eq.langevin}) the momentum diffusion coefficient,
$\kappa$, can be obtained from the correlation of the force term:
\\ 
\beq 
\kappa \ = \ \frac{1}{3} \ \int_{- \infty}^\infty dt \ \sum_i
\langle \xi_i(t) \xi_i(0) \rangle .
\label{eq.force} \eeq
The drag coefficient, $\eta_D$, can be connected to the diffusion
coefficient using standard fluctuation-dissipation relations
\cite{kapusta}: \\ 
\beq 
\eta_D = \frac{\kappa}{2 M T}.
\label{eq.fd} \eeq
Here $M$ is the heavy quark mass.

To have a field theoretic generalization of Eq. (\ref{eq.force}), one
first introduces the conserved current for the heavy quark number
density, $J^\mu(\vec{x}, t) = \bar{\psi} (\vec{x}, t) \gamma^\mu
\psi(\vec{x}, t) $, where $\psi$ is the heavy quark field
operator. The force acting 
on the heavy quark is given by $M \ d J^i/dt$ and so, Eq. (\ref{eq.force})
generalizes to \\
\beq
\kappa \ = \ \frac{1}{3}  \ \lim_{\omega \to 0} \left[ \lim_{M \to
    \infty} \ \frac{M^2}{T \ \chi^{00}} \ \int_{- \infty}^\infty dt \ e^{i \omega
    (t-t^\prime)} \ \int d^3 x \ \left\langle \frac{1}{2} \left\{ \frac{d
  J^i(\vec{x}, t)}{dt} \ , \ \frac{d J^i(\vec{0}, t^\prime)}{dt^\prime} 
\right\} \right\rangle \right] \cdot
\label{eq.kappa1} \eeq
where $\chi^{00}$ is the spatial integral of the density correlator:
\beq
\int d^3 x \langle J^0  (\vec{x}, t) \  J^0  (\vec{0}, t) \rangle \ =
\ T \chi^{00}
\label{eq.susc} \eeq
and is directly proportional to the number density for
a system of non-relativistic quarks.

Since we are working in the heavy quark limit, the force term and the
number density term are easy
to infer: \\
\beqa
M \frac{d J^i}{d t} \ & = & \ \left\{ \phi^\dagger E^i \phi \ -
\ \theta^\dagger E^i \theta \right\} , \nonumber \\
J^0 \ & = & \ \phi^\dagger \phi \ + \ \theta^\dagger \theta 
\label{eq.hq} \eeqa
where $\phi$ and $\theta$ are the two-component heavy quark and
antiquark field operators, respectively, and $E^i$ is the color
electric field. In leading order expansion in $1/M$, only the
electric field contributes to the force term. 

With the substitution of Eq. (\ref{eq.hq}), the real time correlator 
in Eq. (\ref{eq.kappa1}) can be calculated as the analytical continuation
of the Matsubara correlator, \\
\beq
G_E(\tau) \ = \ - \frac{1}{3} \sum_{i=1}^3 \ \lim_{M \to \infty}
\ \frac{1}{T \chi^{00}} \ \int d^3x \ \left\langle  \left\{
\phi^\dagger E^i \phi \ - \ \theta^\dagger E^i \theta \right\}(\tau,
\vec{x}) \   \left\{
\phi^\dagger E^i \phi \ - \ \theta^\dagger E^i \theta \right\}(0,
\vec{0}) \right\rangle \cdot
\label{eq.euclid} \eeq
 
The spectral function, $\rho(\omega)$, for the force term is connected 
to $G_E(\tau)$ by the integral equation \cite{kapusta} \\
\beq
G_E(\tau) \ = \ \int_0^\infty \frac{d \omega}{\pi} \ \rho(\omega)
\ \frac{\cosh \omega (\tau - \frac{1}{2 T})}{\sinh \frac{\omega}{2 T}} \cdot
\label{eq.spectral} \eeq
The momentum diffusion coefficient, Eq. (\ref{eq.kappa1}), is then
given by \\
\beq
\kappa \ = \ \lim_{\omega \to 0} \ \frac{2 T}{\omega} \ \rho(\omega) \cdot
\label{eq.kappa2} \eeq 

Since we are working in the limit of infinitely heavy quarks, the
expression (\ref{eq.euclid}) simplifies considerably. The heavy quark
correlators give a static color field, besides an exponential
suppression factor coming from the heavy quark mass: $\langle 
\theta_a (\tau, \vec{x}) \theta_b^\dagger (0, \vec{0})\rangle = 
\delta^3 (\vec{x}) \ U_{ab}(\tau, 0) \ \exp(-M \tau)$, where 
$U_{ab}(\tau, 0)$ is the timelike gauge connection, and the delta function 
comes because the infinitely heavy quark does not move spatially.
The exponential factor cancels with a similar factor from $\chi^{00}$, 
resulting in a rather simple expression for the infinitely heavy quarks: \\
\beq
G^{\rm Lat}_E(\tau) \ = \ -\frac{1}{3 L} \ \sum_{i=1}^3 \ \left\langle \re \ \tr \ \left[
U(\beta, \tau) \ E_i(\tau, \vec{0}) \ U(\tau,0) \ E_i(0, \vec{0}) \right]
\right\rangle ,
\label{eq.cor} \eeq
where $L = \tr U(\beta, 0)$ is the Polyakov loop. Once again, 
intuitively it is easy to understand Eq. (\ref{eq.cor}): for the 
infinitely heavy quarks, all
that the force-force correlator gives is the correlator of color
electric fields, connected through Wilson lines, and normalized by the
Polyakov loop. 

In order to connect $G^{\rm Lat}_E(\tau)$ measured on the lattice to
physical correlator of electric fields, we need to multiply by a
renormalization factor: \\ \beq \gt = Z(a) G^{\rm
  Lat}_E(\tau) \label{eq.z} \eeq where $Z(a) = Z_E^2$ is the lattice
spacing dependent renormalization factor for the electric field
correlator. A nonperturbative evaluation of the electric field
operator used here is not available. However, the renormalization
factor is expected to be dominated by the self energy correction,
which can be taken into account by a tadpole correction \cite{lm}. In
fact, with other discretizations of the electric field operator it has
been found that the tadpole factor gives a very close approximation to
the nonperturbative renormalization factor \cite{koma}. Here we use
the tadpole factor to renormalize the electric field.

The extraction of $\rw$ from $\gt$ using Eq. (\ref{eq.spectral}) is an
extremely difficult problem. In general, the kernel in
Eq. \ref{eq.spectral} will have zero modes on a discrete lattice,
making the inversion problem ill-defined. In the ideal case, one may
be able to impose some reasonably general conditions on $\rw$ and be
able to make the problem invertible.  However, in situations like
ours, where $\gt$ is measured only on $O(10)$ data points with errors,
the problem of extraction of $\rw$ becomes a completely ill-posed
problem without any further input.
 
For some problems, a Bayesian analysis, with prior information in the
form of perturbative results, has been useful. In general, though,
a stable application of these techniques require both a very large
number of points in the $\tau$ direction and very accurate data for
$\gt$. For the kind of extended objects we are considering, wrapping
the lattice in the Euclidean time direction, it is very difficult to
obtain both together, as the error on the correlators grows with the 
number of points in the $\tau$ direction.
  
Parameterizing $\rw$ in terms of a small number of
parameters, therefore, seems to be a simple way to make the 
inversion problem well-posed. In our case, the leading order 
perturbative form of the spectral function $\sim b \om^3$. Also 
in the $\om \to 0$ regime, we need $\rw \sim \kappa \om$ to get a 
physical value of the diffusion constant using Eq. (\ref{eq.kappa2}). 
The calculation of Ref. \cite{ct} got $\rw = c \om$ for the
$\mathcal{N} = 4$ supersymmetric Yang-Mills theory. 
Motivated by this, we use a simple 
ansatz for the spectral function,
\beq
\rwo \; = \; a \om \, \Theta(\om - \Lambda) \; + \; b \om^3
\label{eq.w} \eeq
We do not include any running in $b$, which is proportional to
$\alpha_s$ in the leading order.  This approach is similar in spirit
to that used in Ref. \cite{electric} to calculate electrical
conductivity. Note that the NLO PT calculation of Eq. (\ref{eq.cor})
leads to a negative value of $\kappa$, and in general, seems to
deviate more from the lattice correlators than the LO result.  So we
use the LO form for the high $\om$ part.

On the other hand, in the calculation with classical lattice gauge
theory \cite{clgt}, the spectral function of the color electric field
was found to have the behavior \\
\[ \rw \; \sim \; c \tanh \frac{\om \beta}{2} \qquad 
{\rm for \ \ } \om a \ll 1. \] So to crosscheck the dependence on our
assumption, we also use a second fit form, \beq \rwt \; = \; c \tanh
\frac{\om \beta}{2} \; \Theta(\om - \Lambda) \; + \; b \om^3.
\label{eq.t} \eeq
In practice, we use these postulated forms for $\rw$ to evaluate
$\gt$, and fit it to the long distance correlation function measured on
the lattice. At large $\om$, of course, this form is not valid, and a
complicated form, that takes into account the effect of the lattice
Brillouin zones, will have to be considered. We tried using the free
lattice spectral function instead of the $\om^3$ term in Eqn.
(\ref{eq.w}). However, that did not improve the fit quality and in
particular, did not seem to capture the very short distance behavior
of the data any better.  So in this work, we restrict ourselves to the
large distance regime in our fits, and expect that in this regime our
simple form will suffice for a first estimate of the diffusion
coefficient.

\section{Numerical Details}
\label{sec.details}
For the lattice evaluation of the correlator $\gt$, we first need to
choose a discretization of the electric field. Following Ref. \cite{clm}
we choose the discretization \\
\[ E_i (\vec{x}, \tau) \ = \ U_i (\vec{x}, \tau) \ U_4 (\vec{x}+\hat{i},
\tau) \ - \ U_4 (\vec{x}, \tau) \ U_i (\vec{x} + \hat{4}) \]
which is a direct latticization of the relation $E_i = [D_0, D_i]$. As
Ref. \cite{clm} suggests, this form of the discretization of the
electric field is expected to be less ultraviolet sensitive than the
more common discretization in terms of the plaquette variable. 

The numerator of Eq. (\ref{eq.cor}) can then be written as
\\
\beqa
G_{E, {\rm num}}^i (\tau) \ &=& \ C^i(\tau + 1) + C^i(\tau - 1) - 2
C^i (\tau) \nonumber \\ 
C^i(\tau) & = & \prod_{x_4=0}^{t-1} U_4(x_4) \, \cdot \, U_i(t) \, \cdot \,
\prod_{x_4=t}^{t+\tau-1} U_4(x_4)  \, \cdot \, U^\dagger_i(t+\tau) \, \cdot \,
\prod_{x_4=t+\tau}^{\beta-1} U_4(x_4)
\label{eq.lcor} \eeqa

\begin{figure}[tpb]
\begin{center}
\scalebox{0.4}{\includegraphics{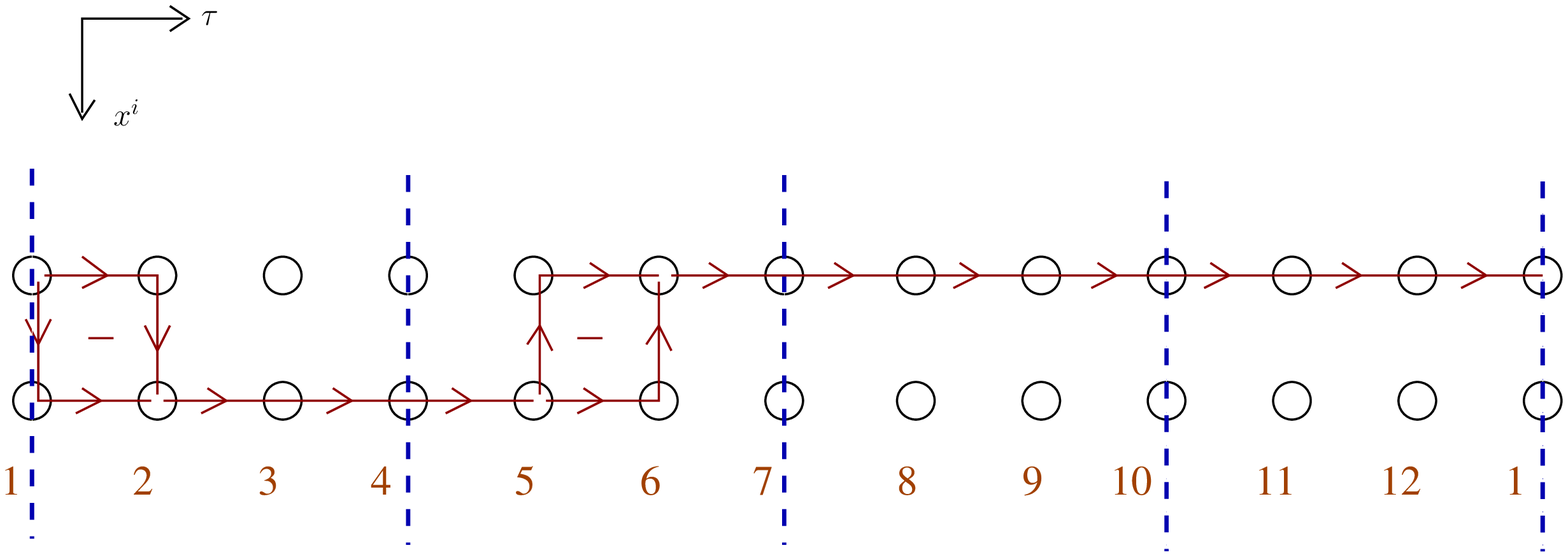}}
\caption{Illustrating the use of the multi-level algorithm for the
calculation of $G^i_{E, {\rm num}}(4)$ on a $N_t=12$ lattice.}
\label{fig.mult}
\end{center}
\end{figure}

The evaluation of $C^i(\tau)$, Eq. (\ref{eq.lcor}), is known to be
difficult for large $\tau$, because the signal-to-noise ratio decays
exponentially. The multilevel algorithm \cite{algo} was indeed devised to
take care of such problems.  We adapted it
for calculation of the electric field correlation functions. The
lattice is divided into several sublattices. The expectation value of
the correlation functions are first calculated in each sublattice by
averaging over a large number of sweeps in that sublattice while
keeping the boundary fixed.  A single measurement is obtained by
multiplying the intermediate expectation values appropriately. The
number of sublattices and the number of sublattice averagings were
tuned for the various sets, so as to get correlators with \% level
accuracy. An explicit example is shown in Fig \ref{fig.mult} which
illustrates the calculation of $G^i_{E,{\rm num}}(4)$ on a $N_t=12$
lattice with four sub-lattices, each with a thickness of three lattice
spacings.  It is important to note that one needs to store all the
intermediate sub-lattice averages separately before they can be
multiplied at end of the update of the whole lattice to construct the
correlation functions. This imposes memory constraints for
simulating large lattices.

The advantage of the multilevel algorithm can be seen from the following
estimate: for $\beta=6.9$, $N_t=20$ and $N_s=36$, the
correlator for $\tau=3a$, $G_E(3)$ has the value of
1.317(2) $\times 10^{-2} $ from 350 multilevel measurements. The
multilevel algorithm takes about 800 minutes to yield a single measurement on an
Intel Xeon CPU processor with a speed of 2.5 GHz.  For the same
correlator, the standard method, using an updating with a combination
of overrelaxation and heatbath steps, led to a value 1.2(2) 
$\times 10^{-2} $ for a runtime of about 8500 minutes on the same
machine. Using the usual $1/\sqrt{t}$ dependence of the error on
runtime, the multilevel algorithm is seen to be about 300 
 times more efficient than the standard algorithm for $G_E(3)$
for this lattice. The efficiency of the multilevel algorithm increases
significantly for larger values of $\tau$. A similar comparison for $G_E(10)$
gives a factor of about 2000 (order of magnitude larger) relative efficiency for the multilevel algorithm. Thus, use of the 
multilevel scheme is indispensable for calculations 
at the larger values of $\tau$ 
\footnote{The efficiency is, of course, dependent on both $\beta$ and $\nt$.},
since these are required to be known with high precision for
the extraction of the diffusion coefficient.

To get the results for various temperatures and volumes, we ran
our simulations at a number of bare couplings with $N_t$ = 12 - 24 and
$N_s/N_t$ = 2 - 4, for temperatures from just above $\tc$ to 3
$\tc$. A reliable extraction of the diffusion coefficient was
possible, however, only for lattices with $\nt \ge 20$. A list of such
lattices used by us is given in Table \ref{tbl.mainlat} below.  To
obtain the temperature scale, we follow the strategy outlined in
\cite{tcscale}. We calculate $\alpha^V$ at each $\beta$ from the
plaquette value. This is translated to a temperature scale at $\nt =
8$ using the information of $\beta_c(\nt = 8)$ \cite{bielefeld} and
two-loop scaling formula with a fitted correction
\cite{ehk}. Temperatures for other $\nt$ are easily calculated from
the $\nt = 8$ temperature scale. The complete list of the lattice
sizes, $\beta$, and the corresponding temperature are shown in Table
\ref{tbl.lattices} in the appendix, which also shows the parameters
used in the multilevel algorithm for each $\beta$.

\begin{table}[htb]
\begin{center} \begin{tabular} {cccccc}
\hline
$\beta$ & 6.76 & 6.80 & 6.90 & 7.192 & 7.255 \\
$\nt$   & 20   & 20   & 20   & 24    & 20    \\
$T/\tc$ & 1.04 & 1.09 & 1.24 & 1.5   & 1.96  \\
\hline
\end{tabular} \end{center}
\caption{List of lattices on which diffusion coefficients were extracted,
and the temperatures.}
\label{tbl.mainlat}
\end{table}

\section{Results}
\label{sec.main}
\begin{figure}[tpb]
\begin{center}
\scalebox{0.7}{\includegraphics{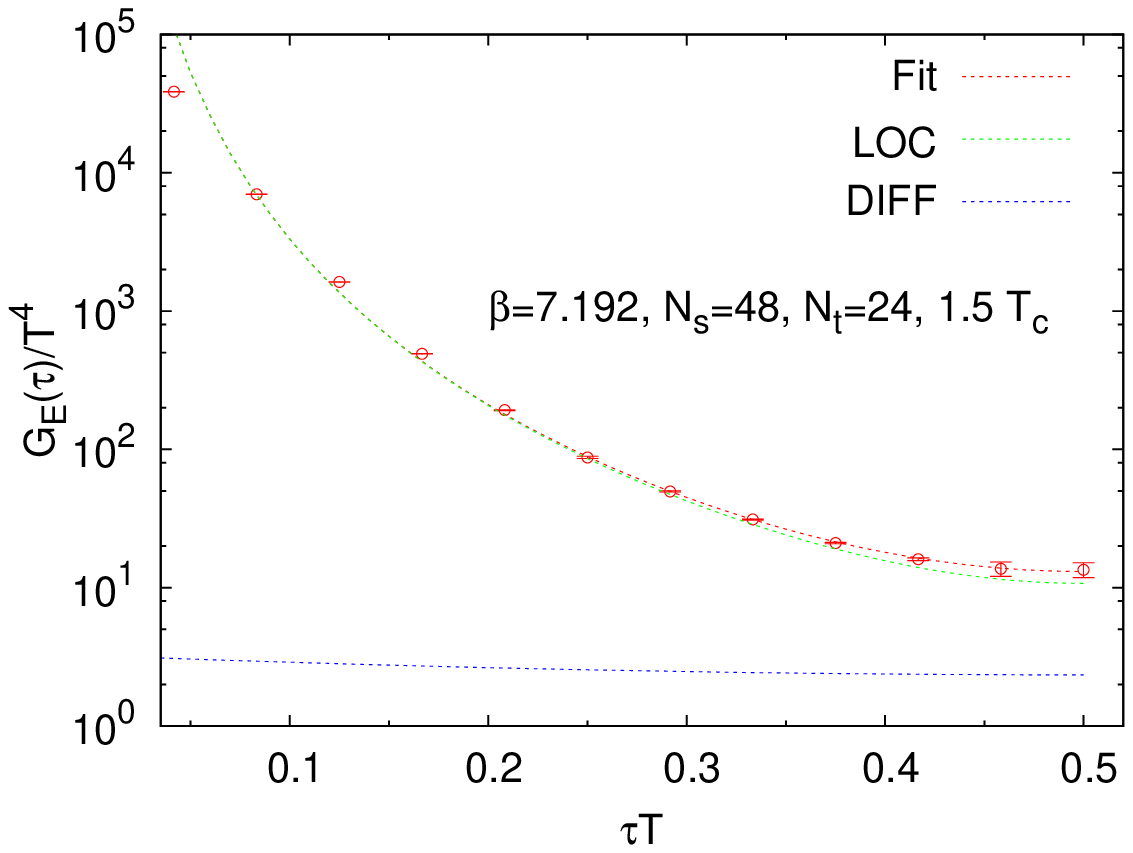}
\includegraphics{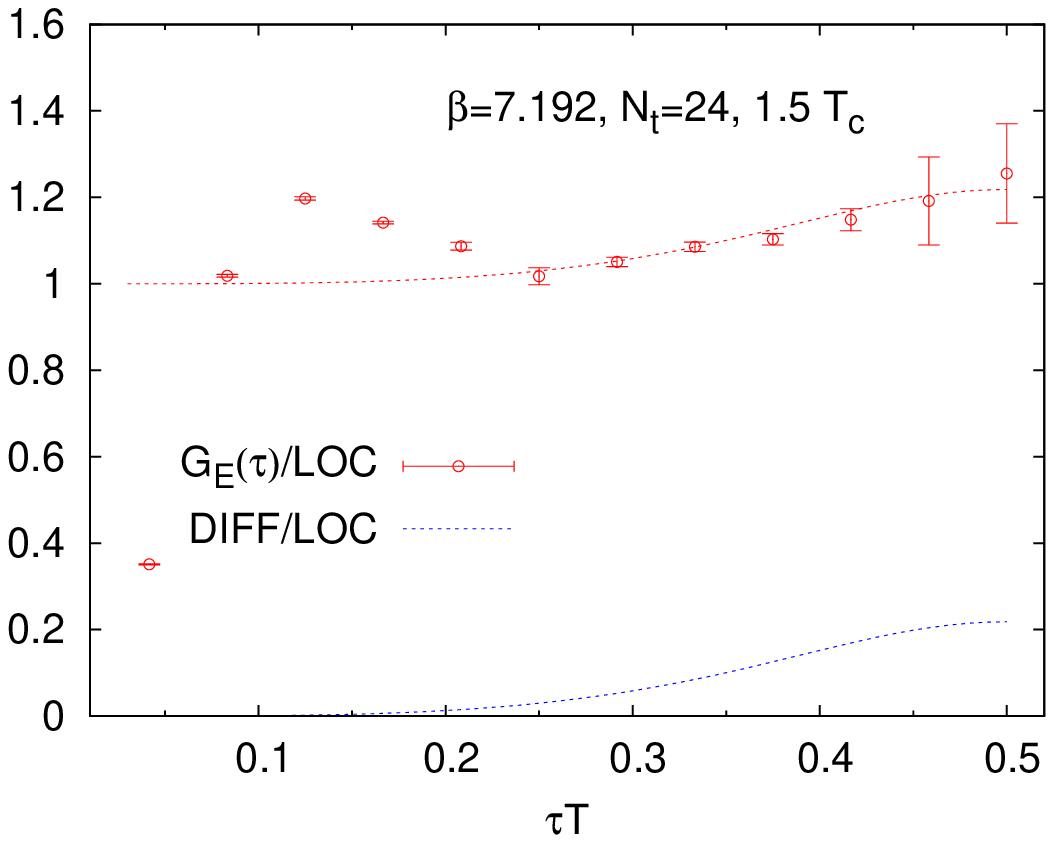}}
\caption{(Left) $\gt$ for one of our lattice sets, at $\beta$ = 7.192 and
  $\nt$ = 24, corresponding to $T = 1.5 \tc$. Also shown is the
  best fit to the form Eq. (\ref{eq.w}) with $\Lambda = 3T$, and contributions
  of the different terms in the fit. LOC corresponds to the correlator
  constructed from the $b \om^3$ term in Eq. (\ref{eq.w}) and DIFF is
  the diffusive part of the correlator, constructed from the first
  term in Eq. (\ref{eq.w}).  (Right) The same information shown
  differently; the measured correlator, the best fit curve, and the
  diffusive part of the correlator are shown normalized to the leading
  order contribution.  }
\label{fig.correl}
\end{center}
\end{figure}

In order to calculate $\kappa$, we calculated the electric field
correlators, Eq.(\ref{eq.cor}), for all the sets in table
\ref{tbl.lattices}.  From the correlators $\gt$, $\kappa$ can be
calculated using Eq. (\ref{eq.kappa2}).  Use of the multilevel
algorithm allowed us to get correlators at a few per cent level accuracy.
In fact, we got $< 2-3$ \% accuracy in all correlators except the two
most central points of the $\beta=7.192, 1.5 \tc$
set. Fig. \ref{fig.correl} shows $\gt$ for this data set.

In order to get the momentum diffusion coefficient, $\kappa$, we use
the ansatz Eq. (\ref{eq.w}) for $\rw$, and fit the Euclidean
correlator using Eq. (\ref{eq.spectral}). It was not feasible to do a
three parameter fit: the parameters $\Lambda$ and $\kappa$ are
strongly correlated. For a large range of $\Lambda$ we can get
very similar fit qualities. Instead, we fix
$\Lambda$ and get an estimate of $\kappa$ by doing a two-parameter
fit. We discuss this further below and in Appendix \ref{sec.syst}.

For the fit, $\chi^2$ minimization was carried out with the full
covariance matrix included in the definition of $\chi^2$.  We
typically obtained acceptable fits to the correlators for $\tau a$ in
the range $[N_t/4, N_t/2]$, with $\chi^2/{\rm d.o.f} \sim 1$.  At shorter
distances, lattice artifacts start contributing and the simple form of
Eq. (\ref{eq.w}) does not work well. Also using the leading order
lattice correlator instead of the continuum form did not improve the
quality of the fit. We, therefore, restrict ourselves to the long
distance part of the correlator.

In order to get a feel for the relative contributions of the different
parts of the spectral function to the correlator, in
Fig. \ref{fig.correl} we show the correlators constructed from
different parts of $\rw$ separately. We take the best
fit form of Eq. (\ref{eq.w}) to the $\nt = 24, 1.5 \tc$ data set, for
$\Lambda = 3 T$. The contributions to the total correlation
function from the $\om^3$ part of $\rw$ and that from the diffusive part, the
first term in Eq. (\ref{eq.w}), are calculated separately using
Eq. (\ref{eq.spectral}). In Fig. \ref{fig.correl} we have called these
parts LOC and DIFF, respectively, and the correlator reconstructed
from the fitted spectral function has been called Fit. The correlator
is seen to be dominated by the contribution from the $b \om^3$ term
over the whole range of distance. However, the diffusion term has a
substantial contribution near the center of the lattice.  In
Fig. \ref{fig.correl} it contributes nearly 20 \% at $\tau T$ =
0.5. This is seen more clearly in the right hand panel of
Fig. \ref{fig.correl}, where the total correlator, the best fit, and
the diffusive part are shown normalized by the leading order
correlator. Since the relative contribution of the diffusive part
falls rapidly at shorter distances, it is difficult to get reliable
estimates of $\kappa$, with the usual assorted tests like stability
with small change in fit range, for our smaller lattices with $\nt$ =
12 and 16. So in what follows, we quote fit results for $\kappa$ only
for our finer lattices, with $\nt =$ 20 and 24 (Table
\ref{tbl.mainlat}).

\begin{figure}[tpb]
\begin{center}
\scalebox{0.7}{\includegraphics{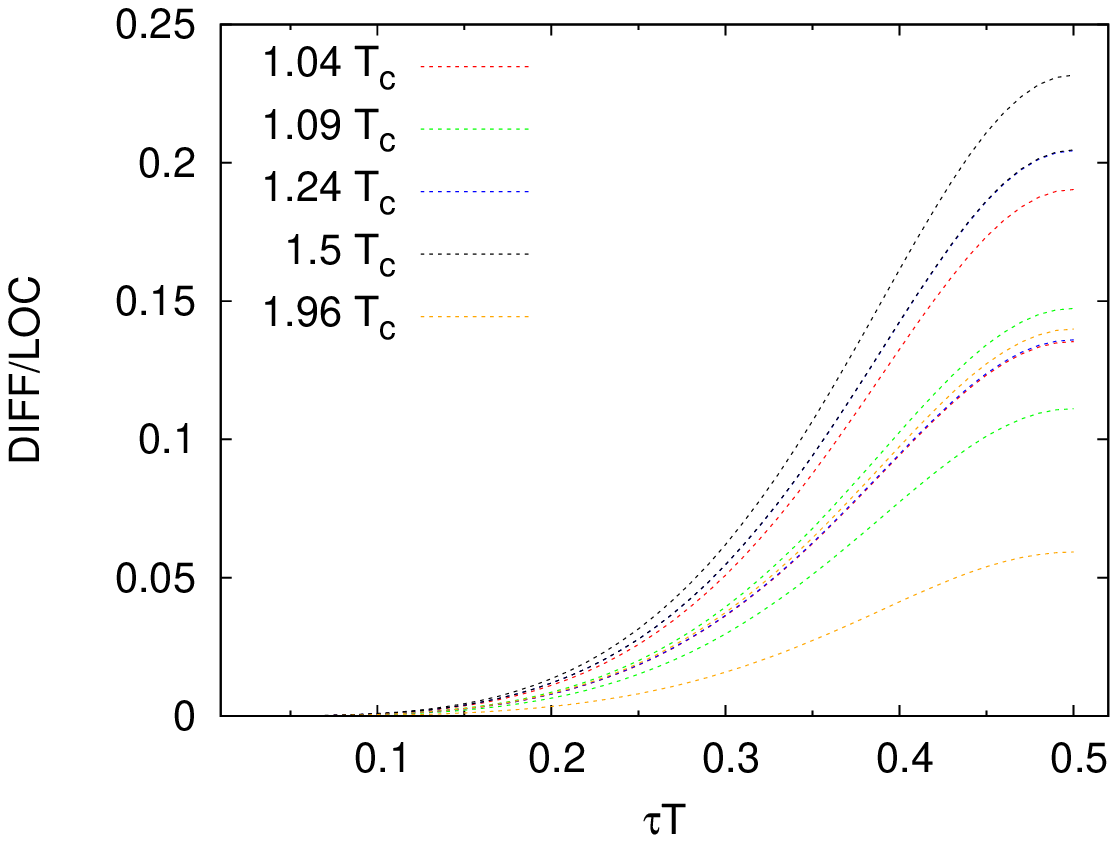}
\includegraphics{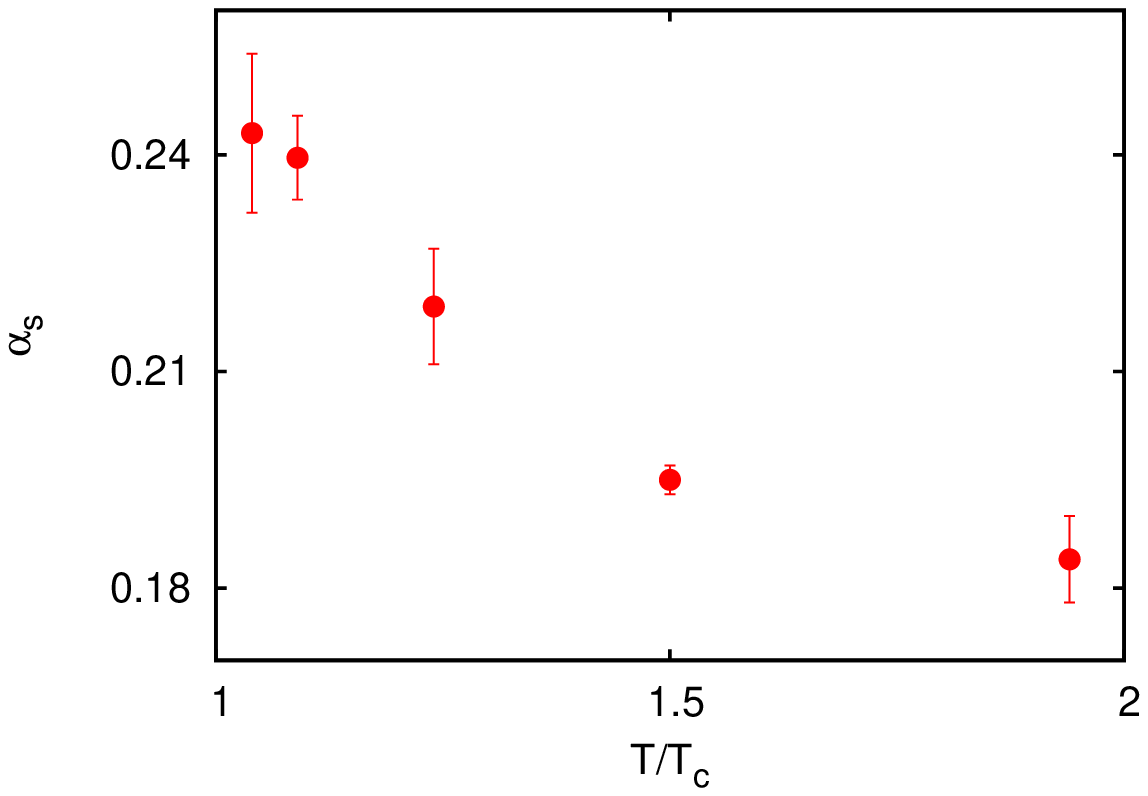}}
\caption{(Left) The relative contribution of the diffusive part(DIFF) to
the total correlator, compared to that of the leading order part (LOC), 
shown as a function of $\tau T$, for our different data sets. 
(Right) $\alpha_s$, defined through the scheme that the coefficient
of the $\om^3$ term in $\rw$ is $8 \alpha_s/9$.}
\label{fig.alphas}
\end{center}
\end{figure}

\begin{figure}[tpb]
\begin{center}
\scalebox{0.9}{\includegraphics{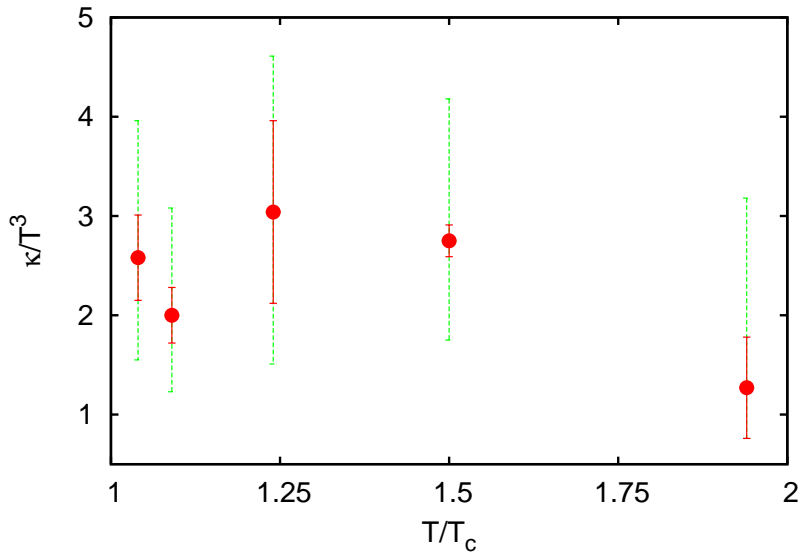}}
\caption{The momentum diffusion coefficient, $\kappa$, in units of
  $T^3$, shown as a function of temperature in the temperature range
  $\tc < T \le 2 \tc$.  The error bars with (red, solid) line show the
  jackknifed error. The (green, dashed) error bars are an estimate of
  the size of the various systematic uncertainties, as discussed in
  the text.}
\label{fig.kp}
\end{center}
\end{figure}

For these lattices, we obtained stable fits for the
central part of the correlator, with $\chi^2 / {\rm d.o.f} \sim 1$ in all
cases. We did a fully correlated fit by including the inverse of the
full covariance matrix in the $\chi^2$ function to be minimized,
whenever such a $\chi^2$ function was well-behaved. That turned out to
be the case in all sets except the one at the highest temperature,
the $\beta = 7.255$ set in Table \ref{tbl.mainlat}. In this case we used 
an uncorrelated fit for our best estimate. The difference between the correlated
and the uncorrelated fit was included in the systematic error.
Our results for $\kappa/T^3$ at various temperatures, using the ansatz
Eq. (\ref{eq.w}) and $\Lambda = 3T$, are shown in Fig. \ref{fig.kp}. The
statistical error, shown by the solid (red) band, is obtained from a
jackknife analysis.  

The choice of $\Lambda = 3T$ for the central value was based on the
fact that in all the sets, with $\Lambda = 3T$, the diffusion term
contribution to the spectral function, $a \om$ in Eq. (\ref{eq.w}), is
numerically close to the large $\om$ term, $b \om^3$, when the
diffusion term sets in (i.e., at $\om = \Lambda$).  Admittedly, this
choice is somewhat arbitrary.  In fact, the main source of uncertainty
in our fit estimate, shown by the dashed (green) band in
Fig. \ref{fig.kp}, is $\Lambda$.  To estimate the possible error
introduced through our central value of $\Lambda = 3T$, we varied
$\Lambda$ in the range $[2 T, \infty)$.  We also looked at the fit
  form Eq. (\ref{eq.t}), and did the same exercise with it. The
  details of the fit results for Eqs. (\ref{eq.w},\ref{eq.t}) and
  various $\Lambda$ are given in Appendix \ref{sec.syst}. Quite often
  the fit value for these variations comes outside the statistical
  error band of Fig. \ref{fig.kp}. A systematic uncertainty band is
  therefore introduced, of sufficient size so as to include the
  central fit values for all these variations.

For the correlation functions of gluonic observables, major finite
volume effects have been observed if the spatial size of the lattice
is so small that some of the spatial directions get deconfined; on the
other hand, at least for spatial correlation functions, finite size
effects are small when the transverse directions are not deconfined
\cite{plb}. To avoid large finite size effects, we choose lattices 
such that the spatial directions are confined.  Since the electric field
correlator also has contribution from the low $\om$ part, it could be more
sensitive to finite volume effects. However, as we discuss in appendix
\ref{sec.syst}, the correlation functions do not show any significant
finite volume effect even when $LT \sim$ 2.  Therefore we do not
expect large finite size corrections to our results obtained from
lattices with $LT \ge 2$.

It is instructive to look at the relative contribution of the
diffusive part to the total correlator at different distances.  In the
left panel of Fig. \ref{fig.alphas} we show the correlator coming from
the diffusive part of the fitted spectral function, normalized by the
leading order part, for all the lattices of Table
\ref{tbl.mainlat}. The notation is similar to that used in
Fig. \ref{fig.correl}, except here we show the 1 $\sigma$ band and not
the best fit value. At all temperatures, except the one at the highest
temperature, the diffusive part is seen to reach about 5 \% level by
$\tau T \sim 0.3$. Note that the accuracy of our correlator is better
than this. Also no significant trend of temperature dependence is seen
in this figure. This is, of course, translated to the lack of
significant temperature dependence of $\kt$ in this temperature
regime, Fig. \ref{fig.kp}.

$b$, the coefficient of the $\om^3$ term in $\rw$, is also of some interest. 
In perturbation theory, the leading order spectral function is \\
\begin{equation}
\rho^{LO}(\om) = \frac{8 \alpha_s}{9} \om^3
\label{eq.alpha} \end{equation}
To get an idea of the strength of the coupling at these temperatures, 
we use Eq. (\ref{eq.alpha}) and the fit coefficient $b$, Eq. (\ref{eq.w}),
for a nonperturbative estimate of $\alpha_s$. 
The estimate of $\alpha_s$ obtained this way is shown in 
Fig. \ref{fig.alphas}. If $\gt$ 
is the properly normalized current, then the NLO calculation of 
Ref. \cite{bllm} can be used to connect this $\alpha_s$ 
to $\alms$. It is interesting to note that the coupling is rather
small, about 1/4 near $\tc$ and going down to $\sim 0.18$ at 2 $\tc$. 
This is in rough agreement with a similar measurement in Ref. \cite{electric}
from fit to vector current correlators, and other, more detailed, calculations 
of $\alpha_s$ at such temperatures from static observables \cite{olaf}.

\begin{figure}[tpb]
\begin{center}
\scalebox{0.9}{\includegraphics{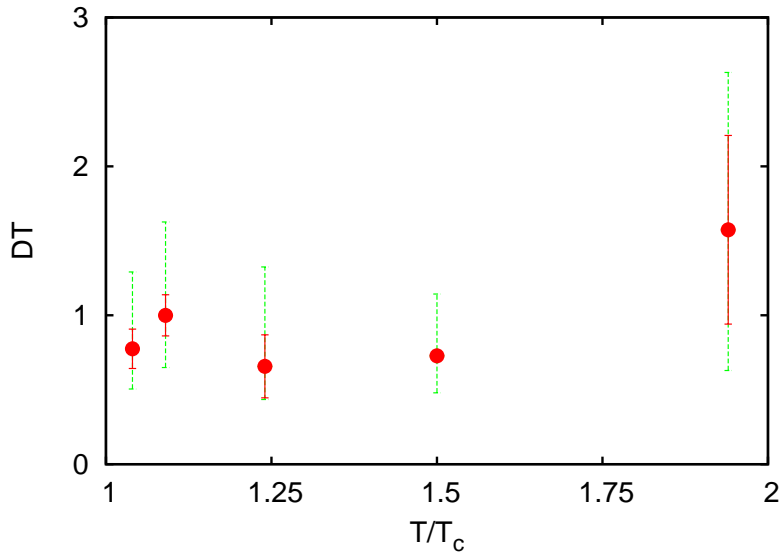}}
\caption{The diffusion coefficient, plotted as $DT$, in the temperature 
range $\sim (\tc, 2 \tc)$. The error bars with solid red lines show the
statistical error. The dashed(green) error bars are an estimate of the 
size of the various systematic uncertainties, as discussed in 
Appendix \ref{sec.syst}.}
\label{fig.dt}
\end{center}
\end{figure}

In order to present our calculation in the context of RHIC, it seems
convenient to use the Einstein relation between the diffusion
coefficient, $D$, and $\kappa$,
\begin{equation}
 D = \frac{T}{M \eta_D} = \frac{2 T^2}{\kappa}.
\label{eq.dt}
\end{equation}
In Eq.(\ref{eq.dt}) $\eta_D$ is the drag constant.  In 
Fig. \ref{fig.dt} we show the
diffusion coefficient in the temperature range $T_c \lesssim T \lesssim 2
\tc$, obtained using Eq. (\ref{eq.dt}). The solid (red) error bar is the 
statistical error from a jackknife analysis.  The bigger
errorbars show the range of values covered by the different
systematics analyzed in Table \ref{tbl.syst}.

Two points are worth noting in this figure. First, the nonperturbative 
value of the diffusion coefficient is rather small in the temperature 
range considered. In the next section we discuss in more detail the comparison
with perturbation theory, but the diffusion coefficient shown here is nearly
an order of magnitude smaller than the leading order perturbation theory. 
Second, there is no strong temperature dependence, at least in the 
temperature range $\tc \le T \le 1.5 \tc$.

\section{Discussion}
\label{sec.summary}
In this work, we studied the momentum diffusion coefficient,
$\kappa$, of heavy quarks in a gluonic plasma. As mentioned in 
the introduction, the large elliptic flow of the heavy flavor mesons, 
seen in the PHENIX experiment at RHIC,
seems to be well explained in a Langevin framework, if $\kappa$ is
large.  Perturbation theory seems to be unstable for this quantity in
the temperature regime of interest for RHIC physics, and the leading
order PT prediction is at least an order of magnitude too small to
explain the experimental data.  Our aim in this work was to
calculate the momentum diffusion coefficient nonperturbatively, and to
see if the deviation from perturbation theory is of the size required
to explain the experimental data.

Using the formalism of Refs. \cite{ct,clm}, we calculated $\kappa$ from
the correlator of $E^a$, the electric field operator. From the
Matsubara correlator of the electric field, $\kappa$ was calculated
through Eq. (\ref{eq.kappa2}) using the ansatz for $\rw$, Eq. (\ref{eq.w}).
In order to compare our results with the
perturbative calculation of \cite{mt} and experiments \cite{exp2}, we
used Eq. (\ref{eq.dt}) to get the diffusion constant, $D$.  The
diffusion coefficient so obtained is found to be considerably 
smaller than the LO PT estimate \cite{mt}. For high temperatures
such that $m_D/T \ll 1$, the leading order estimate of DT is
\cite{mt} \\
\begin{equation}
 DT = \frac{36 \pi}{C_F g^4} \left[ N_c \left( \mathrm{ln}
   \frac{2T}{m_D} + \frac{1}{2} - \gamma_E +
   \frac{\zeta^{\prime}(2)}{\zeta(2)} \right) + \frac{N_f}{2} \left(
   \mathrm{ln} \frac{4T}{m_D} + \frac{1}{2} - \gamma_E +
   \frac{\zeta^{\prime}(2)}{\zeta(2)}\right) \right]^{-1},
\label{eq.lopt}
\end{equation}
where $C_F = (N^2_c-1)/2N_c$ is the color Casimir and $N_f$ is the number
of flavors of thermal quarks. At very high temperatures, $D T$
diverges as $1/\alpha_S^2$. As one comes down in temperature,
Eq. (\ref{eq.lopt}) is not reliable any more and one needs to use the
complete leading order estimate. To get this, we use Eq. (11) of
\cite{mt}, with $\alpha_s(3 T)$ determined using the plaquette
measurement \footnote{The non-perturbative value of $\alpha_S$ at the inverse
  lattice spacing scale, $\mu_1 = (3.4/a) \mathrm{exp}(-5/6)$ is
  obtained from the plaquette values \cite{lm}, and then the 2-loop
  beta function is used to flow to the scale $\mu = 3 T$.} and $m_D$ 
taken from lattice measurements \cite{olaf}. For example, at
1.5 $\tc$, $\alpha_S^{\overline{MS}}(3T) \approx$ 0.23 
and $m_D/T \approx$ 2.345, leading to $D T \simeq 14$. A
similar calculation at 2.25 $\tc$ and 3 $\tc$ yield $DT \simeq$ 18.5
and 21, respectively, for the gluon plasma. A comparison with 
Fig. \ref{fig.dt} reveals that this is almost an order of magnitude larger 
than the nonperturbative result for the gluon plasma.

Interestingly, while Eq. (\ref{eq.lopt}) seems to have a strong dependence 
on $N_f$, on putting values for the different quantities the $N_f=2$ results 
are numerically not very different at similar values of $T/T_c$. 
The next-to-leading order (NLO) contribution to the diffusion constant
has also been calculated in perturbation theory. At similar
temperatures, with $\alpha_S \sim 0.2$, this gives the $DT \sim 8.4/(2
\pi)$ for $N_f=3$ \cite{cm}.  While a similar reduction for $N_f = 0$ 
will bring the NLO PT result much closer to the nonperturbative estimate,
it is rather disconcerting to find that the NLO result differs by almost 
an order of magnitude from the LO result. Indeed, one clearly will have
to resort to calculations of higher orders/resummations before taking
the perturbative estimates seriously.

\begin{figure}[tpb]
\begin{center}
\scalebox{0.7}{\includegraphics{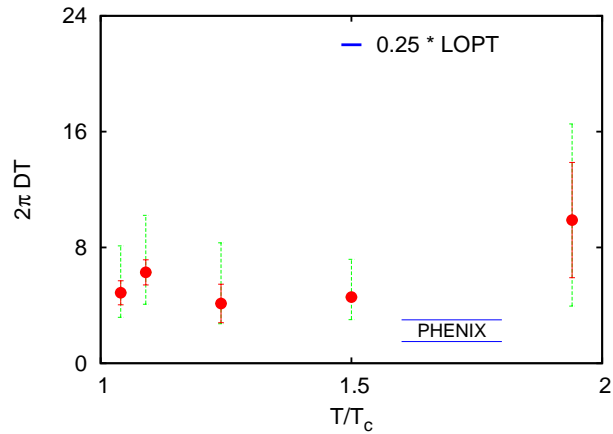}}
\caption{The diffusion coefficient of Fig. \ref{fig.dt}, 
shown here as $2 \pi D T$,
  in the temperature range $\tc < T < 2 \tc$. Also shown is 
the range preferred by the $v_2$ measured by PHENIX \cite{exp2}. 
The band is obtained from a comparison of Fig 40 of Ref.
\cite{exp2} and Fig 4 of \cite{mt}. The LO PT value at 1.5 $\tc$ \cite{mt} 
is also shown.}
\label{fig.exp}
\end{center}
\end{figure}

As already mentioned, there have been other
attempts to calculate the diffusion coefficient using lattice gauge
theory, so far only in the gluon plasma. In Ref. \cite{qm11},
preliminary results for an extraction of the diffusion coefficient
from the vector current correlator $\bar{c} \gamma_i c$ was
presented. The value of $DT$ found at 1.5 $\tc$ was considerably
smaller than LO PT, and is smaller than our results at that temperature, 
though consistent 
within systematics. Ref. \cite{meyer} has also attempted
extracting $\kt$ from the electric field correlator, Eq. (\ref{eq.euclid}),
but with considerably different analysis strategy. This calculation,
which was concentrated mostly on considerably higher temperatures,
found a very small value of $\kt$, which does not agree with ours in
the temperatures where we overlap. On the other hand, a very recent
calculation \cite{langelage}, which also focusses mostly at higher 
temperatures, is in much better agreement with ours in the temperature 
range of overlap.

The heavy quark diffusion coefficient has also been calculated in a very
different theory, the $\mathcal{N} = 4$ supersymmetric Yang-Mills theory 
at large 'tHooft coupling $\lambda_{tH}
= \alpha_S N_c$, using AdS/CFT correspondence \cite{ct,gubser}.  In fact, 
a large part of the formalism used by us 
was introduced in Ref. \cite{ct}. For the $\mathcal{N} = 4$ SYM theory
for $N_c \to \infty$ and large $\lambda_{tH}$, Ref. \cite{ct} gets
\begin{equation}
D T \simeq \frac{0.9}{2 \pi} \left( \frac{1.5}{\lambda_{tH}}
\right)^{\frac{1}{2}}.
\label{eq.ads}
\end{equation}
Note that the dependence of $D$ on the coupling in Eq. (\ref{eq.ads})
is parametrically different from that in Eq. (\ref{eq.lopt}).  Of
course, this theory is very different from QCD in many respect.
Moreover, it exploits crucially symmetries which QCD does not have.
However, to get a feel for what kind of value such a functional
dependence would give, one can somewhat arbitrarily put parameters relevant for
QCD in Eq. (\ref{eq.ads}).  Setting $N_c=3$ and $\alpha_S$ = 0.23, one
obtains $DT \simeq 0.2$ from Eq. (\ref{eq.ads}), which is lower than,
but in the same ballpark as our estimate.

Our results are for quenched QCD, i.e., there are only thermal gluons
but no thermal quarks in our fireball. So a comparison with experimental
results needs to be done with care. A conservative
approach would be to say that comparison of the results in
Fig. \ref{fig.dt} with the perturbative results for quenched QCD give
us an indication of how much the nonperturbative results can change
from the perturbative results in the deconfined plasma at moderate
temperatures $< 2 \tc$. Even then, the
results are most encouraging since they indicate that the
nonperturbative estimate for $DT$ can easily be an order of magnitude
lower than LO PT, bringing it tantalisingly close to values required 
to explain the $v_2$ data. 

In a bit more optimistic fashion, one can hope that our results, as
plotted in Fig. \ref{fig.dt}, will be even quantitatively close to a
similar figure in full QCD when it is computed. The reason for such 
a hope is that dimensionless ratios of various
quantities are known to scale nicely between quenched and full QCD
if plotted as function of $T/T_c$. Also the LO PT result,
Eq.(\ref{eq.lopt}), shows such a trend. In this spirit, in
Fig. \ref{fig.exp} we compare the lattice results with the
experimental data.  The lattice results seem to be a little above the
best fit value for PHENIX, though reasonably close within our large
systematics. Interestingly, our lattice results seem to show very
little temperature dependence in the temperature regime studied
here. For comparison, we also show in the same plot the leading order
PT result for quenched QCD (see estimate below Eq. (\ref{eq.lopt}), 
which is, of course, very far from both the nonperturbative
result and the experimental value. 

The most straightforward direction for possible refinement of our
calculation is, of course, to go to finer and bigger lattices. A
nonperturbative calculation of the renormalization constant will also
be of great help in accurate quantitative prediction.  The nontrivial
next step would be the inclusion of the light thermal quarks in the
calculation. The multilevel algorithm cannot be directly used in that
case, because of the nonlocality of the quark determinant. It would be
an interesting challenge to come up with better ways to obtain
similarly precise results even in the full QCD case.

\section{Acknowledgement}
We would like to thank Sourendu Gupta for numerous discussions, and
for insightful comments on the manuscript.  The computations were done
in the framework of Indian Lattice Gauge Theory Initiative
(ILGTI). The brood cluster of the department of theoretical physics,
TIFR, and the gauge and chiral clusters of the department of
theoretical physics, IACS, were used for this work.  We would like to
thank Ajay Salve and Kapil Ghadiali for technical support.  PM would
like to acknowledge Department of Science and Technology (DST) grant
no. SR/S2/HEP/0035/2008 for the cluster chiral. The research of SD is
partially supported by a Ramanujan fellowship from DST. The research
of RVG is partially supported by a J. C. Bose fellowship from DST.

\appendix
\section{List of lattices, and details of the algorithm}
\label{sec.list}

Here we list the lattices used in our calculations, and parameters for
the multilevel calculation. The parametrs for the multilevel
calculation are also given. The last three columns correspond to the
number of sublattices the lattice was divided in, the number of
sublattice averaging between measurements (\# update), and the number of
measurements of each correlation function (\# conf).

\begin{table}[!tbh]
\begin{center}
\begin{tabular}{c|c|c|c|c|c|c}
\hline
$\beta$ & $N_t$ & $N_s$  & $T/T_c$ & \# sublattice & \# update & \# conf \\
\hline
\multirow{3}{*}{6.4} & \multirow{3}{*}{12}  &  24  &  \multirow{3}{*}{1.07}  
&  3  &  2000  &  190\\
  &  & 36  & & 3  &  2000  &  200\\
& &  48   & &  6  &  200   &  180\\
& & & & & & \\
\multirow{3}{*}{6.65} & \multirow{3}{*}{12} & 24 & \multirow{3}{*}{1.50} &
6  &  200   &  400\\
& & 36 & &  6  &  200   &  260\\
& & 48 & & 6  &  200   &  180\\
& & & & & & \\
\multirow{2}{*}{6.65} & \multirow{2}{*}{16}  &  36  &  \multirow{2}{*}{1.12} &  4  &  2000  &  250\\
& &  48  &   &  4  &  2000  &  215\\
& & & & & & \\
6.76 & 20  &  48  &  1.04  &  5  &  4000  &  170\\
6.80 & 20  &  48  &  1.09  &  5  &  3000  &  150\\
& & & & & & \\
\multirow{6}{*}{6.9} & \multirow{2}{*}{12} & 36 & \multirow{2}{*}{2.07} 
&  6  &  200   &  220\\
& & 48 & &  6  &  200   &  188\\
& \multirow{2}{*}{16} & 36 & \multirow{2}{*}{1.55} 
&  4  &  2000  &  230\\
& & 48 &  & 4  &  2000  &  200\\
& \multirow{2}{*}{20} & 36 & \multirow{2}{*}{1.24}
&  5  &  2000  &  350\\
& & 48 & & 5  &  2000  &  96 \\ 
& & & & & & \\
\multirow{5}{*}{7.192} & 12 & 48 & 3.0 & 3 & 2000 &  210\\
& 16 & 48 & 2.25 & 4  &  2000  &  200\\
& \multirow{3}{*}{24} & 48 & \multirow{3}{*}{1.5} & 4  &  2000  &  450\\
& & 56 & & 4  &  2000  &   50\\
& & 56 & & 4  &  4000  &   45\\
& & & & & & \\
7.255 & 20 & 48 & 1.96 & 5  &  2000  &  194\\
7.457 & 16  &  48 &  3.0    &  4  &  2000  &  140\\ 
\hline
\end{tabular}
\end{center}
\caption{Details of the lattices used in the calculation.
Also given are the parameters for the multilevel algorithm
for each set.}
\label{tbl.lattices}
\end{table}

\section{Details of various systematics discussed in Sec. \ref{sec.main}}
\label{sec.syst}
In this section we discuss in some detail some of the systematic uncertainties mentioned in Sec. \ref{sec.main}. 

\begin{itemize}
\item{\bf Uncertainties in the ansatz for the spectral function:} \\
In Sec. \ref{sec.theory} we have discussed the ansatz for spectral
function used by us to get the diffusion coefficient. As we have
discussed there, we have no first principle handle on the form, and
have used forms for the low-$\om$ part motivated by other studies. In
particular, we have introduced a cutoff in Eqs. (\ref{eq.w},
\ref{eq.t}). As we discussed in Sec.  \ref{sec.main}, the quality of
the fit is rather insensitive to $\Lambda$: as we vary $\Lambda$, we
get a different best fit value for $\kappa$, but for a range of
$\Lambda$, the $\chi^2$ does not change appreciably. This is probably
an example of the zero mode solutions we discussed in
Sec. \ref{sec.theory}. We illustrate this in Fig.  \ref{fig.lambda},
for the ($\beta = 7.192, 1.5 \tc$) set. In the figure the contribution
of the diffusive part of the correlator is shown for the best fit
parameters at various $\Lambda$.  The corresponding values of
$\kappa$, shown in Table \ref{tbl.syst}, vary substantially as
$\Lambda$ is varied; however, the total contribution of the diffusive
part to the correlator hardly changes as we change $\Lambda$ from 2T
to 4T.

\begin{figure}[tpb]
\begin{center}
\scalebox{0.7}{\includegraphics{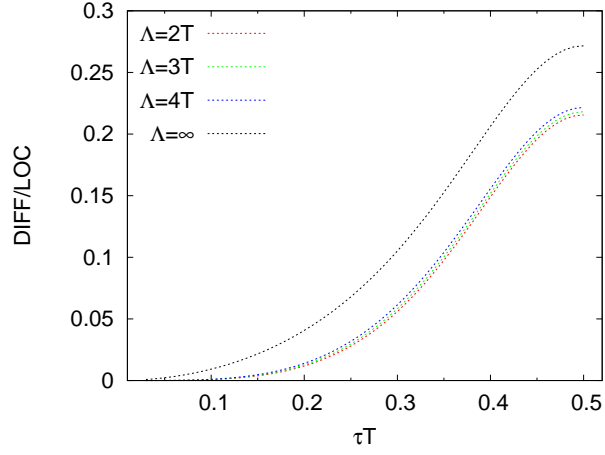}}
\caption{The change in the relative contribution of the diffusive part
of the $\beta = 7.192$, 1.5 $\tc$ data set, as we change $\Lambda$ in Eq. 
(\ref{eq.w}). The notation is same as Fig. \ref{fig.correl}.} 
\label{fig.lambda}
\end{center}
\end{figure}

Of course, the cutoffs in Eqs. (\ref{eq.w},\ref{eq.t}) are an
approximation: one does not expect discreet jumps in $\rw$. It is not
unreasonable to expect, however, that changing the sharp cutoff with a
smooth one will not change things significantly and that the flat
direction we encounter is of more general origin. For the purpose of
this work, we take the conservative approach of letting $\Lambda$ vary
between $[2 T, \infty)$, and include the values of $\kappa$ thus
  obtained in the systematic uncertainty band.  We consider this range
  to be conservative because for $\Lambda < 2 T$, plugging back the
  fit solution to construct $\rw$, we get a large jump at $\om =
  \Lambda$, since at this value $a \om$ in Eq. (\ref{eq.w}) is much
  bigger than $b \om^3$. In order to quote a central value for the
  fits, we investigated for what value of $\Lambda$ $a \om \sim b
  \om^3$ for $\om = \Lambda$. For all the sets of Table
  \ref{tbl.mainlat}, this happens around $\Lambda \sim 3T$. Therefore,
  we use this value of $\Lambda$ to quote the central
  value. Admittedly, this criterion is arbitrary, and the green band
  in Figs. \ref{fig.kp}, \ref{fig.dt} is probably the more robust
  object.
 
In Table \ref{tbl.syst} we also repeat this exercise of varying $\Lambda$ 
for the fit form $\rwt$. When the fit values obtained are outside the 
systematic band, the band is extended to include them. 

\begin{table} [tbh]
\begin{center}
\begin{tabular}{cccc|cccc|cccc}
\hline
$T/T_c$ & $\beta$ & $N_t$ & $N_s$ & \multicolumn{4}{c|}{$\rwo$} 
& \multicolumn{4}{c}{$\rwt$} \\
\hline
& & & & \multicolumn{4}{c|}{$\Lambda = $} & \multicolumn{4}{c}{$\Lambda = $} \\
& & & & $2 T$ & $3 T$ & $4 T$ & $\infty$ & $2 T$ & $3 T$ & $4 T$ 
& $\infty$ \\
\hline
1.04 & 6.76 & 20 & 48 & $3.6 \pm 0.6$ & $2.6 \pm 0.4$ & $2.1 \pm 0.4$ 
& $1.55 \pm 0.26$ & $4.0 \pm 0.7$ & $3.0 \pm 0.5 $ & $2.6 \pm 0.4$ & 
$2.2 \pm 0.4$ \\
1.09 & 6.80 & 20 & 48 & $2.8 \pm 0.4$ & $2.0 \pm 0.3$ & $1.6 \pm 0.2$ 
& $1.23 \pm 0.17$ & $3.1 \pm 0.4$ & $2.4 \pm 0.3$ & $2.06 \pm 0.29$ & 
$1.77 \pm 0.24$ \\
1.24 & 6.90 & 20 & 48 & $3.5 \pm 0.7$ & $2.5 \pm 0.5$ & $2.0 \pm 0.4$ & 
$1.5 \pm 0.3$ & $3.8 \pm 0.8$ & $2.9 \pm 0.6$ & $2.5 \pm 0.5$ & 
$2.2 \pm 0.4$ \\
    &      &    & 36 & $3.5 \pm 0.6$ & $2.5 \pm 0.4$ & $2.1 \pm 0.3$ &
$1.5 \pm 0.3$ & $3.8 \pm 0.6$ & $2.9 \pm 0.5$ & $2.6 \pm 0.4$ & $2.2 \pm 0.4$\\
1.50 & 7.192 & 24 & 48 & $3.8 \pm 0.2$ & $2.75 \pm 0.16$ & $2.22 \pm 0.13$ & 
$1.75 \pm 0.10$
& $4.18 \pm 0.24$ & $3.19 \pm 0.18$  & $2.80 \pm 0.16$ & $2.45 \pm 0.14$ \\ 
1.96 & 7.255 & 20 & 48 & $1.8 \pm 0.7$ & $1.3 \pm 0.5$ & $1.0 \pm 0.4$ 
& $0.81 \pm 0.33$ & 
$1.9 \pm 0.8$ & $1.5 \pm 0.6$ & $1.3 \pm 0.5$ & $1.14 \pm 0.46$ \\
\hline
\end{tabular}
\label{tbl.syst}
\caption{Fit form dependence of $\kappa/T^3$.}
\end{center} \end{table}

\item{\bf Fit range and fit quality dependence:}
For the fit values quoted in Table \ref{tbl.syst}, we have used the
range $\tau_{\rm min}$ to $\nt /2$, where $\tau_{\rm min}$ is the
smallest $\tau$ for which we got a good $\chi^2$, and the $\chi^2$ is
defined using the full covariance matrix. In all sets except one, we
could get a good $\chi^2$ with $\tau_{\rm min} \ge \nt/4$, and
increasing $\tau_{\rm min}$ slightly did not change
the fit value significantly. The set where we could not get such a
stability with $\tau_{\rm min}$ was the set at $\beta$ = 7.255. In
this case only the uncorrelated $\chi^2$, i.e., the diagonal
covariance matrix in the definition of $\chi^2$, allowed such
stability. So for this set, we used the uncorrelated fit value for our
central estimate. The difference between the uncorrelated and the
correlated best fits is then taken as an additional source of
systematic uncertainty in this case. In fact, the systematic
uncertainty band for this set in Fig. \ref{fig.kp} is dominated by
this contribution.

\item{\bf Finite volume effect:}
We explored finite volume effects by looking at $LT$ =
2-4 on some of our coarser lattices, and smaller variations of LT in two of
our finer lattices. At the correlation function level itself, no
statistically significant finite volume effect could be seen once $LT \ge 2$. 
To make this statement quantitative, we do a $\chi^2$ comparison of
the long distance part of the correlator, which should be the most sensitive
to finite volume effects. For the correlator calculated on two
lattices at the same $\beta$ and $\nt$ but different $N_s$, we
construct the quantity
\[ \chi^2/d.o.f. = \frac{1}{\nt /4} \ 
\sum_{\tau=\frac{\nt}{4}+1}^{\nt /2} \frac{\vert 
G_1(\tau)-G_2(\tau) \vert}{\sqrt{\sigma_1(\tau)^2+\sigma_2(\tau)^2}} \cdot \]
For the different sets in Table \ref{tbl.lattices} this
quantity is listed below. \\
\begin{center} \begin{tabular}{cccc}
\hline
$\beta$ & $N_t$ & ($LT\vert_1,  LT\vert_2$) & $\chi^2$/d.o.f. \\
\hline
6.4 & 12 & (2, 4) & 0.34 \\
\multirow{2}{*}{6.65} & 12 & (2, 4) & 0.75 \\
     & 16 & (2.25, 3) & 1.12 \\
\multirow{3}{*}{6.9} & 12 & (3, 4) & 0.24 \\
      & 16 & (2.25, 3) & 0.51 \\
      & 20 & (1.8, 2.4) & 1.58 \\
7.192 & 24 & (2, 2.33) & 0.29 \\
\hline
\end{tabular} \end{center}
Here the third column shows the $LT$ values of the lattices 
whose correlators are being compared.
At the level of accuracy of our correlators, we do not see any
significant finite volume effect for LT = 2. So we believe our
results, calculated on admittedly small lattices, will not be severely
affected by finite volume effects.  Even for the LT=1.8 set at $\beta$
= 6.9, $\nt = 20$, where the correlator does show a statistically
significant effect, Table \ref{tbl.syst} reveals that the error in
$\kappa$ due to finite volume effect is smaller than our other systematics.

\item{\bf Renormalization factor:} \\
The lattice correlator is multiplied by a renormalization factor,
Eq. (\ref{eq.z}), to get $\gt$. Clearly, an error in $Z_E$ will affect
$\kappa$ multiplicatively. As we discussed in Sec. \ref{sec.theory},
in the absence of a nonperturbative evaluation of the renormalization
factor, we have used the tadpole factor, which takes into account the
quadratic self energy correction of the gluon lines, to renormalize
$G_E^{\rm Lat}$.

A perturbative renormalization factor, using heavy quark effective
theory, has been calculated in Ref. \cite{langelage}. Below we
tabulate the two renormalization factors for the different lattice
spacings we have: \\
\begin{center} \begin{tabular}{l|lllll}
\hline 
$\beta$ = & 6.76 & 6.80 & 6.9 & 7.192 & 7.255 \\ 
$T/T_c$ = & 1.04 & 1.09 & 1.24 & 1.5 & 1.96 \\ 
\hline
$Z^{\rm tad}$ & 1.230 & 1.232 & 1.226 & 1.210 & 1.207 \\ 
$Z^{\rm HQET}$ & 0.831 & 0.832 & 0.834 & 0.841 & 0.842 \\ 
\hline
\end{tabular} \end{center} 
Over the temperature range of interest to us, there is a near-constant
factor 1.43-1.48 between the two renormalization schemes. We do not
include such a factor in our band of systematics, since it is easy to
convert our results to $Z^{\rm HQET}$ factor. We note that while this
indicates a rather large reduction in $\kappa$ of order 30-32 \%, it
will not change our qualitative conclusions.

\end{itemize}

\end{document}